%% file: draft_conference_v6.tex
\documentclass[conference,10pt]{IEEEtran}

\usepackage{comment,color,cite,url}
\usepackage{subfigure,times,graphicx}
\usepackage{amsmath}
\usepackage{amssymb}

\def\BibTeX{{\rm B\kern-.05em{\sc i\kern-.025em b}\kern-.08em
    T\kern-.1667em\lower.7ex\hbox{E}\kern-.125emX}}

\DeclareMathAlphabet{\mathsfbf}{OT1}{cmss}{sbc}{n}

\newcommand{\PP}{\mathbb{P}} 
\newcommand{\RR}{\mathbb{R}} 

\newcommand{\ee}{{\rm e}}

\newcommand{\ev}{{\bf e}}

\newcommand{\qv}{{\bf q}}
\newcommand{\rv}{{\bf r}}
\newcommand{\sv}{{\bf s}}

\newcommand{\uv}{{\bf u}}
\newcommand{\wv}{{\bf w}}
\newcommand{\vv}{{\bf v}}
\newcommand{\xv}{{\bf x}}
\newcommand{\yv}{{\bf y}}
\newcommand{\zv}{{\bf z}}

\newcommand{\Am}{{\bf A}}
\newcommand{\Bm}{{\bf B}}
\newcommand{\Cm}{{\bf C}}

\newcommand{\Hm}{{\bf H}}
\newcommand{\Id}{{\bf I}}

\newcommand{\Km}{{\bf K}}

\newcommand{\Pm}{{\bf P}}

\newcommand{\Zm}{{\bf Z}}

\newcommand{\Dc}{{\cal D}}
\newcommand{\Ec}{{\cal E}}

\newcommand{\Nc}{{\cal N}}

\def\ben{\begin{enumerate}}
\def\beq{\begin{equation}}
\def\beqa{\begin{eqnarray}}
\def\bit{\begin{itemize}}
\def\een{\end{enumerate}}
\def\eeq{\end{equation}}
\def\eeqa{\end{eqnarray}}
\def\eit{\end{itemize}}

\newcommand{\ls}[1]
   {\dimen0=\fontdimen6\the\font
    \lineskip=#1\dimen0
    \advance\lineskip.5\fontdimen5\the\font
    \advance\lineskip-\dimen0
    \lineskiplimit=.9\lineskip
    \baselineskip=\lineskip
    \advance\baselineskip\dimen0
    \normallineskip\lineskip
    \normallineskiplimit\lineskiplimit
    \normalbaselineskip\baselineskip
    \ignorespaces
   }

\title{Anytime Reliable LDPC Convolutional Codes for Networked Control over Wireless Channel}
\author{Alberto Tarable, Alessandro Nordio, Fabrizio Dabbene, Roberto Tempo}

\author{
\IEEEauthorblockN{Alberto Tarable, Alessandro Nordio, Fabrizio Dabbene, and Roberto Tempo
}
\IEEEauthorblockA{CNR-IEIIT, Torino, Italy\\
\{alberto.tarable, alessandro.nordio, fabrizio.dabbene, roberto.tempo\}@ieiit.cnr.it}  
}

\begin{document}
\maketitle

\begin{abstract}
This paper deals with the problem of stabilizing an unstable system through networked control over
the wireless medium. In such a situation a remote sensor communicates the measurements to the system
controller through a noisy channel. In particular, in the AWGN scenario, we show that
protograph-based LDPC convolutional codes achieve anytime reliability and we also derive a lower
bound to the signal-to-noise ratio required to stabilize the system. Moreover, on the
Rayleigh-fading channel, we show by simulations that resorting to multiple sensors allows to achieve
a diversity gain.
\end{abstract}

\section{Introduction}

In the field of control theory, it is of growing interest the study of networked control systems,
where the measurement sensor and the controller are not physically co-located. Such a case is suitably
modelled by supposing that the remote sensor transmits its measurements to the controller through a noisy
communication channel. From the information-theoretical point-of-view, such communication problem
has many differences from the ordinary reliability problem on a point-to-point link. Such
differences arise essentially from the fact that systems must be controlled in real-time, while the
usual approach does not consider delay as a primary parameter. Moreover, past decoding errors at the
receiver may have a catastrophic effect if they are not eventually corrected as time proceeds.

Based on the previous considerations, Sahai and Mitter~\cite{Sahai} introduced the new concepts of
\emph{anytime reliability} and of \emph{anytime capacity}. Loosely speaking, an encoding-decoding
scheme is said to be anytime reliable if its bit error probability decreases exponentially with
delay $d$, i.e., is proportional to\footnote{In this paper, we use the Euler number $\ee$ instead of
  2 as the base of the exponential.} $\ee^{-\beta d}$, where $\beta$ is the \emph{anytime exponent}
of the scheme. Then, the anytime capacity $C(\beta)$ is the supremum of achievable rates for schemes
with anytime exponent $\beta$. For further information, see~\cite{Sahai_th} and references therein.

Anytime-reliable nonlinear tree codes were first proven to exist in \cite{Schulman} and then further
developed in \cite{Ostrovsky}. Random linear codes were first introduced in \cite{Como}. Later,
Sukhavasi and Hassibi \cite{Hassibi} showed that causal random linear codes with maximum-likelihood
(ML) decoding are anytime reliable with high probability. Such schemes are characterized by a high
decoder complexity, although in \cite{Hassibi} a decoder with reasonable complexity is proposed for
the erasure channel.  In~\cite{Dossel} Dossel et al. proposed a low-density parity-check (LDPC)
convolutional encoding scheme on the erasure channel which is shown to be anytime reliable. In this
case, a belief-propagation decoder allows for achieving anytime reliability at an affordable
complexity.

Since the erasure channel is more suitable for modeling the behavior of upper layer communications,
in this work, we study anytime reliable LDPC convolutional codes over the wireless channel. The
rationale behind this choice is that in this case LDPC decoding algorithms can make use of soft
information and thus are expected to be more efficient. Moreover, a physical-layer encoding-decoding
scheme can better exploit the potentialities of a wireless multi-node control network where a
network of remote sensors transmit simultaneously their measurements to the system controller.

\begin{figure*}[tb]
\centerline{\resizebox{0.7\textwidth}{!}{
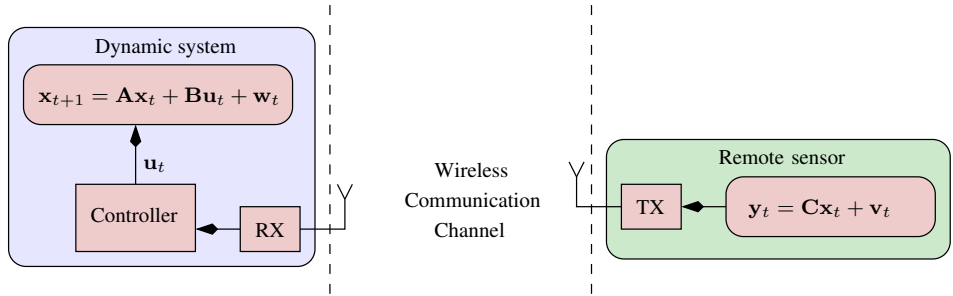}}
\caption{Model of the discrete-time linear dynamic system:
  remote sensor, wireless communication channel and system controller.}
\label{fig:model}
\end{figure*}

The contributions of the paper are as follows:
\begin{itemize}
\item We generalize the design of the LDPC convolutional codes proposed by~\cite{Dossel} to a
  broader class.
\item We prove that such class of codes are anytime reliable on the AWGN channel, and we give a
  lower bound on their anytime exponent.
\item We show by simulations that, in the presence of fading channel between the sensor and the
  controller, better stability margins can be achieved by using multiple sensors.
\end{itemize}

The paper is organized as follows. In Section \ref{sec:model}, we describe the (single-sensor) model
of the considered system. In Section \ref{sec:encoder_decoder}, we generalize the LDPC convolutional
codes introduced in \cite{Dossel}. In Section \ref{sec:anytime}, we derive a lower bound on the
anytime exponent for the LDPC convolutional encoding scheme on the AWGN channel. Eventually in
Section \ref{sec:results}, through numerical analysis, we validate the theoretical results obtained
in the previous sections and we show simulation results for the sensor network scenario.

\section{System model}
\label{sec:model}
We consider the discrete-time dynamic time invariant system\footnote{ Column vectors and matrices
  are denoted by lowercase and uppercase bold letters, respectively.}
\begin{equation}
\xv_{t+1} = \Am \xv_t + \Bm\uv_t + \wv_t 
\label{eq:system}
\end{equation}
where $\xv_t\in \RR^{n_x}$ is the state of the system at time step $t$, $\Am$ and $\Bm$ are $n_x
\times n_x$ and $n_x \times n_u$ real matrices, respectively, $\uv_t$ is the control input, and
$\wv_t$ is a zero-mean bounded noise process. The system in~\eqref{eq:system} is supposed to be
unstable, i.e., it is characterized by $\rho(\Am)>1$, where $\rho(\Am)$ is the spectral radius of
the matrix $\Am$, that is the largest eigenvalue modulus of $\Am$.  The state $\xv_t$ of the linear
system in~\eqref{eq:system} is measured by a remote sensor (see Figure~\ref{fig:model}) providing
the size-$n_y$ measurement
\begin{equation} \label{eq:sensor}
\yv_t = \Cm\xv_t + \vv_t
\end{equation}
where $\vv_t \in \RR^{n_y}$ is a zero-mean bounded noise process independent of
$\wv_t$. 

The remote sensor is equipped with one antenna and sends its measurement to a controller through a
noisy wireless communication channel. Specifically, in the considered setting the sensor at each
time step $t$ first quantizes the measurement $\yv_t$ into a $k$-bit vector $\qv_t$. In this paper,
as in \cite{Hassibi}, we consider a uniform lattice quantizer. The quantized measurements are the
input of a channel encoder $\Ec_t$, whose output $\ev_t$ is a binary vector of length $n$. The code
rate is thus $R_c=k/n$.  The encoded bits are modulated to a vector of $m$ symbols, $\sv_t$,
belonging to a given constellation and then trasmitted over the wireless channel.

The received signal is given by
\begin{equation}
 \rv_t = \Hm_t\sv_t +\zv_t
\label{eq:Recevied_signal}
\end{equation}
where $\zv_t$ is a size-$m$ vector representing additive noise with i.i.d. complex circularly
symmetric Gaussian random entries with zero mean and unitary variance. The $m\times m$ diagonal
channel matrix $\Hm_t=\mathrm{diag}(h_{t1},\ldots,h_{tm})$ is such that its $j$-th diagonal element
represents the channel coefficient experienced at time $t$ by the $j$-th transmitted symbol, for
$j=1,\dots,m $.

At time step $t$ the receiver processes the received signal $\rv_t$ and obtains the soft estimates
of the quantized measurements.  Specifically, $\rv_t$ is first processed by a ML demodulator which
outputs soft estimates $\hat{\ev}_t$ of the coded bits $\ev_t$.  The soft estimates $\hat{\ev}_t$
and the estimates computed at time steps $\tau=1,\ldots,t-1$, i.e., $\hat{\ev}_1, \ldots,
\hat{\ev}_{t-1}$ are then sent to a decoder $\Dc_t$ which outputs estimates
$\hat{\qv}_{0|t},\ldots,\hat{\qv}_{t|t} $ of the quantized measurements.  The notation
$\hat{\qv}_{\tau|t}$ represents the estimate of $\qv_{\tau}$ obtained at the receiver at time step
$t$.

Finally a digital-to-analog converter provides estimates $\hat{\yv}_{1|t},\ldots, \hat{\yv}_{t|t}$
of the observations. Again the notation $\hat{\yv}_{\tau|t}$ represents the estimate of $\yv_{\tau}$
obtained at time step $t$.  These estimates are sent to the system controller.

The system controller is in charge of generating suitable commands $\uv_t$ in order to keep the
system stable. In particular at time $t$ the controller takes as input the estimates
$\hat{\yv}_{1|t},\ldots, \hat{\yv}_{t|t}$, and outputs the control $\uv_t$.  The controller is a
chain of $t$ filters. The $\tau$-th filter produces the output $\hat{\xv}_{\tau|t}$, which is an
estimate of the state $\xv_\tau$ at time step $t$, and has two inputs: the estimate
$\hat{\xv}_{\tau-1|t}$ and the vector $\hat{\yv}_{\tau|t}$.  The output of the $t$-th filter,
$\hat{\xv}_{t|t}$ is then providing the estimate of the current state.  The command $\uv_t$ is
finally obtained as a linear feedback of the state estimate $\hat{\xv}_{t|t}$
\[ \uv_t = \Km\hat{\xv}_{t|t}\]
where the $n_x\times n_u$ matrix $\Km$ is chosen to stabilize the system, i.e.\ so that
$\rho(\Am+\Bm\Km)<1$.  In this paper, hypercuboidal filters~\cite{Hassibi} have been employed in the
system controller.

\section{LDPC convolutional encoding/decoding schemes}
\label{sec:encoder_decoder}

Following \cite{Dossel}, we consider a channel coding scheme based on systematic LDPC convolutional
codes.  Precisely, the encoder at time $t$, $\Ec_t$, is the systematic encoder corresponding to the
parity-check matrix \beq \Zm_{[1:t]} = \left[
\begin{array}{ccccc}
\Zm_0 & & & & \\
\Zm_1 & \Zm_0 & & & \\
\vdots & \Zm_1 & \ddots & & \\
\vdots & \vdots & \ddots & \ddots & \\
\Zm_{t-1} & \Zm_{t-2} & \dots &  \Zm_1 & \Zm_0
\end{array}\right]
\eeq
 where all matrices $\Zm_i$, $i=0,\dots,t-1$, are $(n-k) \times n$ sparse binary matrices, and
$\Zm_0$ is full-rank (over $\mathbb{GF}(2)$) in order for $\Zm_{[1:t]}$ to have full row
rank. (Notice that we have explicitly restricted our focus to a \emph{Toeplitz} parity-check matrix,
for simplicity.) Such coding scheme is causal thanks to the lower-triangular structure of the
parity-check matrix. Moreover, it can be considered as a convolutional code with infinite memory,
whose number of states is equal to $2^{(t-1)k}$ at time $t$.

Structurally, as in \cite{Dossel}, we have built the parity-check matrix $\Zm_{[1:t]}$ starting from
a protograph \cite{Thorpe} matrix $\Pm_{[1:t]}$, given by \beq \Pm_{[1:t]} = \left[
\begin{array}{ccccc}
\Pm_0 & & & & \\
\Pm_1 & \Pm_0 & & & \\
\vdots & \Pm_1 & \ddots & & \\
\vdots & \vdots & \ddots & \ddots & \\
\Pm_{t-1} & \Pm_{t-2} & \dots &  \Pm_1 & \Pm_0
\end{array}\right]
\eeq where all matrices $\Pm_i$, $i=0,\dots,t-1$, are $(n_0-k_0) \times n_0$ matrices with
nonnegative integer entries and $\Pm_0$ is full-rank. $\Zm_{[1:t]}$ is obtained by \emph{lifting}
$\Pm_{[1:t]}$ to order $r$, namely:
\begin{itemize}
\item each zero of $\Pm_{[1:t]}$ is lifted to a $r \times r$ all-zero matrix, and
\item each nonzero entry of $\Pm_{[1:t]}$ equal to $b$ is lifted to the modulo-2 sum of $b$
  permutation matrices of size $r \times r$, chosen at random between the $r!$ possible permutation
  matrices of that size.
\end{itemize} 
In this way, we obtain a $(n-k)t \times nt$ sparse binary matrix $\Zm_{[1:t]}$, with $(n-k) = r
(n_0-k_0)$ and $n = r n_0$. Notice that $R_c = k/n = k_0/n_0$. Since permutation matrices are chosen
at random, we actually obtain an \emph{ensemble} of codes, each one corresponding to a given choice
of the permutations. The fact that $\Zm_{[1:t]}$ is sparse even if $\Pm_{[1:t]}$ is not, allows us
to choose the latter with a certain degree of freedom. In particular in our work we choose
\beq \label{eq:B0} \Pm_0 = \left[ \Pm_{0,p} \Big| \Id_{\overline{k}_0} \right] \eeq and, for
$\tau=1,2,\dots$ \beq \label{eq:Btau} \Pm_{\tau} = \left[ \Pm_{\tau,p} \Big|
  \mathbf{0}_{\overline{k}_0 \times \overline{k}_0} \right] \eeq where $\overline{k_0} =
n_0-k_0$. Note that the choice of the protograph matrices in~\eqref{eq:B0} and~\eqref{eq:Btau} is
more general with respect to the choice made in~\cite{Dossel} where $\overline{k_0}=1$ and $n_0=2$.

In our system the decoder $\Dc_t$ implements belief propagation (BP)~\cite{Richardson} on the bipartite graph defined
by $\Zm_{[1:t]}$. In this case the advantage of deriving the code by lifting a protograph relies on
the fact that the \emph{local} structure of the code graph always looks like the protograph one~(\cite{Thorpe, Liva}), while the probability of short cycles (detrimental for
BP) is reduced by increasing $r$. Thus, convergence properties of BP decoding algorithm can be
studied directly on $\Pm_{[1:t]}$, while neglecting the effects of cycles.

As for ordinary LDPC codes~\cite{Richardson}, $\mathbf{P}_{[1:t]}$ is interpreted as the adjacency matrix of the protograph at time $t$, where columns represents variable nodes (VNs) and rows represent check nodes (CNs). If a given element of $\mathbf{P}_{[1:t]}$ is equal to $b$, there are $b$ edges connecting the corresponding CN and VN. Moreover, if $b>0$, the two nodes are neighbors. Let $\Nc_c(i)$ be the set of VNs that are neighbours of $i$-th
CN. Analogously, let $\Nc_v(j)$ be the set of CNs that are neighbours of VN $j$.

\section{Anytime reliability of LDPC convolutional codes on the AWGN wireless channel}
\label{sec:anytime}
In this section, we study the anytime reliability of the protograph-based LDPC convolutional codes.
We first provide a theoretical analysis where we derive a bound on the bit error probability of the
LDPC encoding/decoding scheme for a generic noisy channel. Then in Section~\ref{sec:AWGN} we
specialize to the AWGN case.
  
\subsection{A lower bound on the anytime exponent of LDPC encoding/decoding schemes}
Let $P_e(t,d)$ be the probability that, at time $t$, the oldest decoding error made by the decoder
$\Dc_t$ is $d$ steps back in the past: 
\beq \label{eq:Petd_def}
\begin{split}
P_e(t,d) = \PP \{& \{\hat{\qv}_{t-d+1|t}\neq\qv_{t-d+1}\} \cap  \\
 &\{\hat{\qv}_{\tau|t}=\qv_\tau, \tau < t-d+1\}\}. 
\end{split}
\eeq We say that the encoding-decoding scheme is \emph{anytime reliable} on a given channel if it
satisfies: 
\beq 
P_e(t,d) <K \ee^{-\beta d},\,\,\, \forall t, \,d>d_0 \label{eq:Petd} 
\eeq 
where $K$, $\beta$ and $d_0$ are positive constants that depend on the coding scheme and on the
channel. If the code satisfies (\ref{eq:Petd}), then $\beta$ is called its \emph{anytime exponent}
(on that channel).  

Sukhavasi and Hassibi derive in \cite{Hassibi} the conditions under which an anytime reliable
encoding-decoding scheme can be used to stabilize the system of (\ref{eq:system})-(\ref{eq:sensor})
in the mean-square sense, so that the expected value of $\|\mathbf{x}_t\|^2$ is bounded for all $t$.
It is shown in \cite{Hassibi} that, when using hypercuboidal filters, mean-square sense stability is
achieved by a code with anytime exponent $\beta$ satisfying $\beta > 2\log\rho(\overline{\Am})$,
where $\overline{\Am}$ is the $n_x \times n_x$ matrix whose elements are the absolute values of the
elements of $\Am$.

 In order to assess the performance of the BP decoder, in what follows, we slightly modify the P-EXIT approach of \cite{Liva,Dossel}. In particular,
we suppose that the BP messages exchanged between VNs and CNs
are sent through AWGN channels, and
we track the evolution of the SNR of such channels with the iteration index of the BP algorithm.

Let us define the following variables: 
\begin{itemize}
\item $\rho_{ch}(j)$: the physical-channel SNR for VN $j$.
\item $\rho_{C \rightarrow V,t}^{(l)}(i,j) $: the SNR for message travelling from CN $i$ to 
VN $j$ at the $l$-th iteration of the BP algorithm and at time step $t$ (if $i \in \Nc_v(j)$).
\item $\rho_{V \rightarrow C,t}^{(l)}(i,j) $: the SNR for message travelling from VN $j$ to CN $i$
  at the $l$-th iteration of the BP algorithm and at time step $t$ (if $i \in \Nc_v(j)$).
\end{itemize}

Then, the approximate SNR evolution at time $t$ can be determined through the following set of
update equations\footnote{For simplicity, the update equations are given in the hypothesis that the
  protograph matrix is binary. However, the results hold in general. }~\cite{Tarable}:
\begin{itemize}
\item {\bf Initialization:} For $j=1,\dots,n_0 t$, $i \in \Nc_v(j)$:
\beq \label{eq:initialization}
\rho_{V \rightarrow C,t}^{(0)}(i,j) = \rho_{ch}(j)
\eeq
\item {\bf CN to VN update:} For $i=1,\dots,\overline{k}_0 t$, $j\in \Nc_c(i)$: 
\begin{equation} \label{eq:CNupdate}
\rho_{C \rightarrow V,t}^{(l+1)}(i,j) \simeq M\left( \sum_{\substack{s\in \Nc_c(i)\\ s \neq j}}
M \left(\rho_{V \rightarrow C,t}^{(l)}(i,s)\right)\right)
\end{equation}

\item {\bf VN to CN update:} For $j=1,\dots,n_0 t$, $i \in \Nc_v(j)$:
\begin{equation} \label{eq:VNupdate}
\rho_{V \rightarrow C,t}^{(l+1)}(i,j) = \sum_{\substack{s \in \Nc_v(j) \\ s \neq i}}
\rho_{C \rightarrow V,t}^{(l+1)}(s,j)+ \rho_{ch}(j)
\end{equation}

\item {\bf Output decision variable SNR:} For $j=1,\dots,n_0 t$:
\begin{equation} \label{eq:outputSNR}
\rho_{t}^{(l+1)}(j) = \sum_{s \in \Nc_v(j)}
\rho_{C \rightarrow V,t}^{(l+1)}(s,j)+ \rho_{ch}(j)
\end{equation}
\end{itemize}

The function $M(\rho)$ appearing in (\ref{eq:CNupdate}) is defined as $M(\rho) = J^{-1} \left( 1-J(\rho) \right)$, where
\beq 
J(\rho) = 1 - \int_{-\infty}^{+\infty} \frac{e^{-(y-2\rho)^2/(8\rho)}}{\sqrt{8 \pi \rho}}
\log_2 (1+e^{-y}) dy
\eeq
gives the mutual information between the input of a binary-input AWGN channel with SNR $\rho$ and the corresponding output.
Notice that $M(\rho)$ is a nonnegative, strictly decreasing function of $\rho$ and that
$M^{-1}(\rho) = M(\rho)$.  

%

It can be easily proven~\cite{Tarable} that the sequences $\rho_{C \rightarrow V,t}^{(l)}(i,j)$ and
$\rho_{C \rightarrow V,t}^{(l)}(i,j)$ are monotonically increasing with iteration index
$l$. Moreover, these sequences are bounded, as long as $\rho_{ch}(j)$ is bounded for all $j$. Thus,
they converge to a limit when $l$ goes to infinity. Let us call such limits $\rho_{C \rightarrow
  V,t}^{(\infty)}(i,j)$ and $\rho_{V \rightarrow C,t}^{(\infty)}(i,j)$, which are a function of the
channel SNR values.  The output decision variable SNR for VN $j$, after a large number of
iterations, is then given by \beq \label{eq:rhot_infinity} \rho_{t}^{(\infty)}(j) = \sum_{s \in
  \Nc_v(j) } \rho_{C \rightarrow V,t}^{(\infty)}(s,j)+ \rho_{ch}(j).  \eeq

Next, starting from (\ref{eq:Petd_def}) we compute an upper bound to $P_e(t,d)$ after a large number of BP iterations as follows:
\begin{eqnarray}
P_e(t,d) & \stackrel{(a)}{\leq} & \PP \{\hat{\qv}_{t-d+1|t}\neq\qv_{t-d+1}\}  \nonumber \\
& \stackrel{(b)}{\leq} & \sum_{i=1}^{k_0} Q\left(\sqrt{\rho_{t}^{(\infty)}((t-d) n_0 + i)}\right) \nonumber \\
& \stackrel{(c)}{\leq} & \frac1{2}\sum_{i=1}^{k_0} \ee^{-\rho_{t}^{(\infty)}((t-d) n_0 + i)/2} \nonumber \\
& \leq & \frac{k_0}{2} \ee^{-\min_{i=1}^{k_0} \rho_{t}^{(\infty)}((t-d) n_0 + i)/2} \label{eq:PetdUB2}
\end{eqnarray}
where (a) follows from (\ref{eq:Petd_def}), (b) follows from the fact that the code is systematic and the union bound and (c) from the Chernoff bound on $Q(x)$. Thus, thanks to (\ref{eq:Petd}) and (\ref{eq:PetdUB2}), the anytime exponent of the considered coding scheme can be lower-bounded
by 
\beq \label{eq:underline_beta_def} \beta \geq \underline{\beta} = \lim_{d \rightarrow
  \infty}\frac{\min_{i=1}^{k_0} \rho_{t}^{(\infty)}((t-d) n_0 + i)}{2d}.  
\eeq 

Thus, a sufficient condition for the stabilization of system (\ref{eq:system})-(\ref{eq:sensor})
in the single-node scenario is that $\underline{\beta} > 2\log\rho(\overline{\Am})$. The value of
$\underline{\beta}$ can be obtained numerically, for a given coding scheme, thanks to
(\ref{eq:initialization})-(\ref{eq:outputSNR}).

\subsection{The AWGN case}
\label{sec:AWGN}
For the AWGN case, $\rho_{ch}(j) = \rho_{ch} $ for all $j$. Notice that, because of (\ref{eq:B0})
and (\ref{eq:Btau}), every systematic VN is connected to a CN that is connected to a degree-1
nonsystematic VN. Since the message coming from a degree-1 VN is set to $\rho_{ch}$ at every
iteration, thanks to the fact that $M(\rho)$ is monotonically decreasing with $\rho$, we can
upper-bound the CN-to-VN message exchanged at iteration $l$ as 
\beq M\left( \sum_{\substack{s\in
    \Nc_c(i)\\ s \neq j}} M (\rho_{V \rightarrow C,t}^{(l)}(i,s))\right) \leq \rho_{ch} \eeq which
clearly holds also for $l \rightarrow \infty$. By plugging the above bound into
(\ref{eq:rhot_infinity}), we can write the following upper bound on the output SNR for systematic
variable $j$: \beq \label{eq:outputSNRUB} \rho_{t}^{(\infty)}(j) \leq (|\Nc_v(j)|+1) \rho_{ch}.
\eeq It is shown in \cite{Tarable} that this upper bound is actually reached for $\rho_{ch}
\rightarrow \infty$ if some mild conditions are satisfied.

Thus, for sufficiently large $\rho_{ch}$, the lower bound on the anytime exponent can be approximated by
\beq
\underline{\beta} = \lim_{d \rightarrow \infty}\frac{\min_{i=1}^{k_0} (|\Nc_v((t-d) n_0 + i)|+1) \rho_{ch}}{2d}.
\eeq

In order to achieve anytime reliability for sufficiently large $\rho_{ch}$, the VN degrees must then
increase linearly with $d$. If $ \min_{i=1}^{k_0} |\Nc_v((t-d) n_0 + i)| = \gamma d + o(d)$ for all $t$, then
\beq \label{eq:beta_bar_awgn}
\underline{\beta} = \frac{\gamma \rho_{ch}}{2}.
\eeq
The above results tells that, in order to stabilize the system in (\ref{eq:system})-(\ref{eq:sensor}) over an AWGN channel with the LDPC convolutional encoding-decoding scheme and hypercuboidal filters, it is sufficient that the channel SNR satisfies $\rho_{ch} > 4  \log\rho(\overline{\Am})/ \gamma$.

Notice also that, from the analysis, it seems beneficial to increase $\gamma$, which corresponds to using a denser protograph: however, while this is true for a lifting order $r$ going to infinity, increasing $\gamma$ may affect negatively the performance for a finite $r$, due to the increased probability of finding short cycles. 

\section{Results}
\label{sec:results}
To validate the theoretical results, we have simulated a system characterized by
\beq
\mathbf{A} = \left[ 
\begin{array}{ccc}
1.285 & 0.127 & 0. \\
4 & 1.285 & 0.002 \\
-3.94 & -0.280 & 0.979
\end{array}
 \right] \eeq with $\rho(\overline{\Am}) = 1.997$, and with the matrices $\Bm$ and $\Cm$ chosen as
in \cite[Example 1]{Hassibi}.
Moreover we used the LDPC convolutional code of \cite{Dossel}, with
$k_0=1$, $n_0=2$ and $\Pm_{\tau,p} = 1$ for all $\tau$, so that $\gamma=1$.

\begin{figure}[!t]
    \centerline{
    \includegraphics[width=0.9\columnwidth,angle=0]{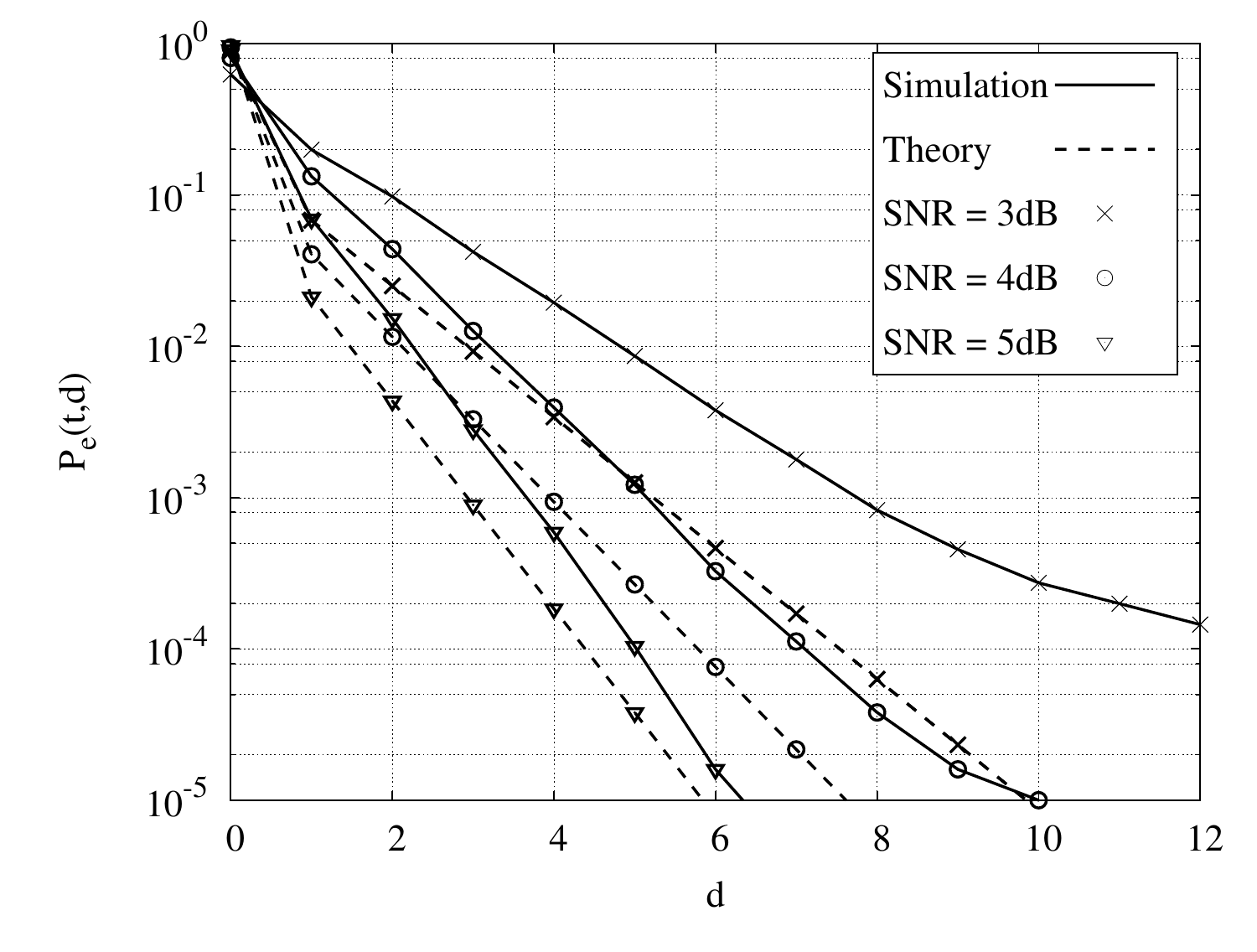}}
   \caption{Anytime exponent of the LDPC convolutional codes on the AWGN channel: theory versus simulations.}
\label{Fig:res_AWGN}
\end{figure}

If the channel between sensor and controller can be modelled as AWGN, then, by using
(\ref{eq:beta_bar_awgn}), we conclude that the system can be stabilized in the mean-square sense as
long as $\rho_{ch} > 4 \log \rho(\overline{\Am}) = 4.4$ dB. Fig.~\ref{Fig:res_AWGN} shows the error
probability $P_e(t,d)$ versus the delay $d$.  The experimental curves (solid lines) obtained by
simulations, are compared with the theoretical ones (dashed), obtained according
to~(\ref{eq:PetdUB2}), in the hypothesis that the upper bound of~(\ref{eq:outputSNRUB}) is actually
achieved. As it can be seen, for $\rho_{ch} > 4$ dB, the slope of the simulated $P_e (t,d)$, which
corresponds to the anytime exponent, is larger than the theoretical one, as predicted from the
analysis. Monte Carlo simulations actually show that the system is controlled for $\rho_{ch} = 4.5$
dB.

\subsection{Extension to the fading scenario}
\label{sect:fading}

Through simulations, we have investigated an extension of the control system of Section
\ref{sec:model} to the fading scenario in the case of a network made of multiple sensors. More
precisely, we consider the case where there are $N$ identical remote sensors, whose measurements are
subject to independent bounded noise.  At time $t$, the $i$-th sensor, $i = 1,\dots,N$, obtains the
size-$n_y$ measurement 
\beq \yv_t^{(i)} = \Cm\xv_t + \vv_t^{(i)} \eeq 
where $\vv_t^{(i)}$, $i=1,\dots,N$, are zero-mean bounded noise processes independent of each other 
and of $\wv_t$.  
The $i$-th sensor encodes the information as described in Section \ref{sec:encoder_decoder} and
transmits the symbol vector $\sv_t^{(i)}$ to the common receiver, which is equipped with $N$
antennas. The signal received at the $j$-th receive antenna, $j = 1,\dots,N$, will then by given by
\beq \rv_t^{(j)} = \sum_{i=1}^N \Hm_t^{(j,i)}\sv_t^{(i)} +\zv_t^{(j)} \eeq where the diagonal
channel matrix $\Hm_t^{(j,i)}$ contains on its diagonal the channel coefficients from sensor $i$
and receive antenna $j$. We assume Rayleigh fading, so that the instantaneous SNR at the receiver is
a random variable exponentially distributed as
\beq f_{\rho_{ch}}(\rho) = \frac{1}{\overline{\rho}_{ch}} \ee^{-\rho/\overline{\rho}_{ch}} \eeq

\begin{figure}[!t]
    \centerline{
    \includegraphics[width=0.9\columnwidth,angle=0]{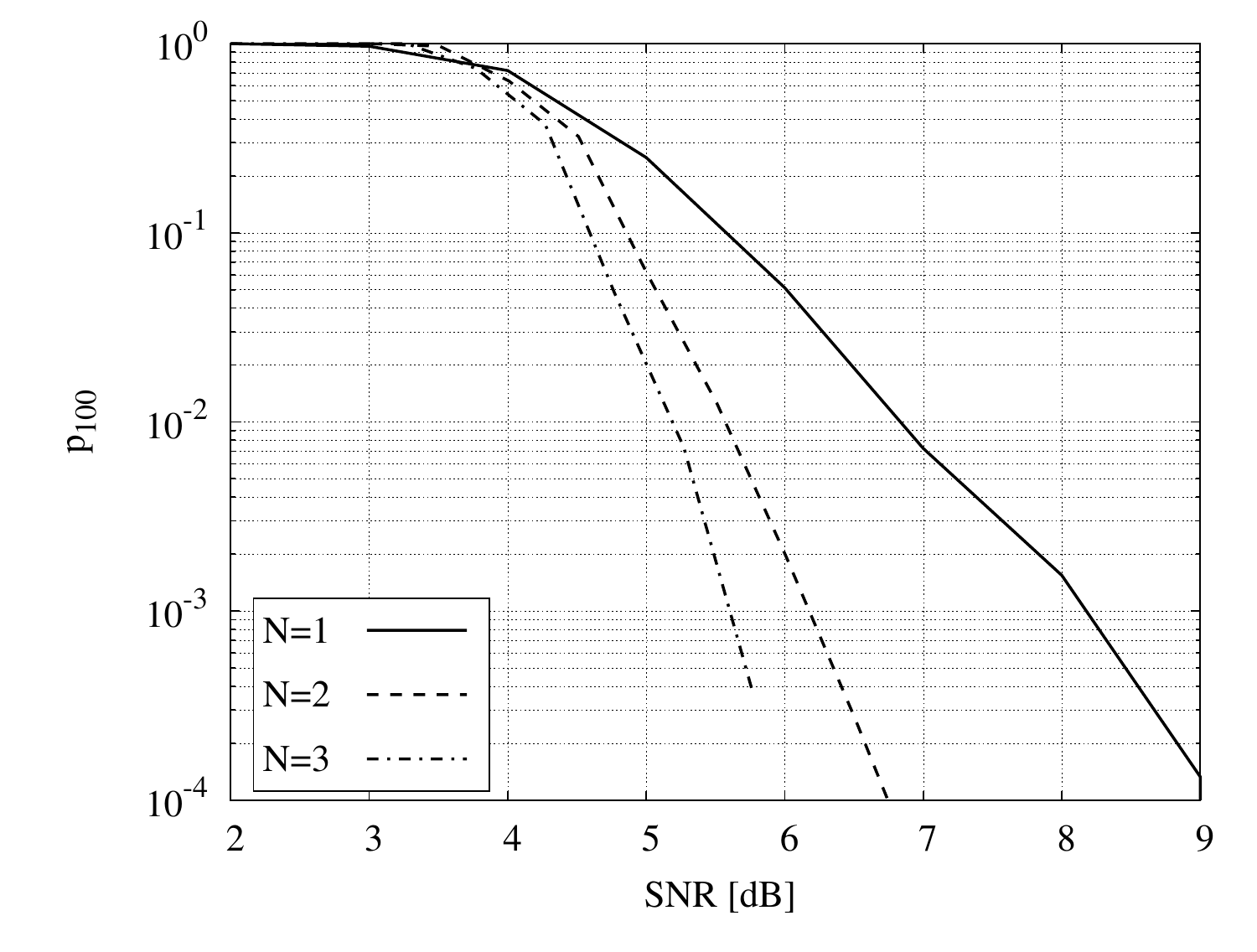}}
   \caption{Performance of multiple sensors in the fading scenario.}
   \label{Fig:res_fading}
\end{figure}

The receiver performs jointly optimal demodulation of the $N$ superimposed transmitted
signals. After demodulation, $N$ decoders work in parallel to decode the information sent by the $N$
sensors. The controller computes the feedback signal $\uv_t$ by putting together the reconstructed
measurements from all sensors.
In Fig. \ref{Fig:res_fading}, we show the performance of the simulated control network. We have
measured the probability, $p_{100}$, that, after 100 time steps, the Euclidean distance between the
system state and the estimated state is larger than $10^3$, as a function of the average SNR,
$\overline{\rho}_{ch}$, for $N=1,2,3$. The power transmitted by each sensor has been normalized so
that the total transmitted power is the same for the three cases. We have used a lifting order
$r=12$ in the LDPC code construction. As it can be seen, $p_{100}$ decreases with
$\overline{\rho}_{ch}$. Notice also that the slope of $p_{100}$ increases with
$N$. We can conclude that a diversity gain can be obtained by using  multiple sensors.

\bibliographystyle{IEEE}

\end{document}

%% file: system_1observer.pdf_t
\begin{picture}(0,0)%
\includegraphics{system_1observer.pdf}%
\end{picture}%
\setlength{\unitlength}{4144sp}%
\begingroup\makeatletter\ifx\SetFigFont\undefined%
\gdef\SetFigFont#1#2#3#4#5{%
  \reset@font\fontsize{#1}{#2pt}%
  \fontfamily{#3}\fontseries{#4}\fontshape{#5}%
  \selectfont}%
\fi\endgroup%
\begin{picture}(5694,1779)(259,-2818)
\put(1756,-2446){\makebox(0,0)[lb]{\smash{{\SetFigFont{8}{9.6}{\rmdefault}{\mddefault}{\updefault}{\color[rgb]{0,0,0}RX}%
}}}}
\put(1081,-2041){\makebox(0,0)[lb]{\smash{{\SetFigFont{8}{9.6}{\rmdefault}{\mddefault}{\updefault}{\color[rgb]{0,0,0}$\uv_t$}%
}}}}
\put(451,-1636){\makebox(0,0)[lb]{\smash{{\SetFigFont{8}{9.6}{\rmdefault}{\mddefault}{\updefault}{\color[rgb]{0,0,0}$\xv_{t+1} = \Am \xv_t + \Bm\uv_t + \wv_t $}%
}}}}
\put(4051,-2311){\makebox(0,0)[lb]{\smash{{\SetFigFont{8}{9.6}{\rmdefault}{\mddefault}{\updefault}{\color[rgb]{0,0,0}TX}%
}}}}
\put(4726,-2311){\makebox(0,0)[lb]{\smash{{\SetFigFont{8}{9.6}{\rmdefault}{\mddefault}{\updefault}{\color[rgb]{0,0,0}$\yv_t = \Cm\xv_t + \vv_t$}%
}}}}
\put(4546,-2011){\makebox(0,0)[lb]{\smash{{\SetFigFont{8}{9.6}{\rmdefault}{\mddefault}{\updefault}{\color[rgb]{0,0,0}Remote sensor}%
}}}}
\end{picture}%